# Design, construction, commissioning and long term operational experience with the D0 Uranium/Liquid Argon calorimeter

**Dean Schamberger**[*]
*Stony Brook University*
*E-mail:* Dean.Schamberger@StonyBrook.edu

The D0 experiment was designed in the mid 1980s and ran at the Fermilab $p\overline{p}$ collider from 1992 through 2011. I describe the uranium-liquid argon calorimeter and its readout electronic which was upgraded in the late 1990s to handle the higher luminosity of the upgraded Tevatron during its second running period from 2001-2011. I summarize maintaining the calorimeter for 20 years of data taking. I further describe a few issues that arose during that time, including different types of noise and the anomalous high voltage currents seen only in the central calorimeter.

[*]Speaker.

## 1. D0 Calorimeter

The D0 Uranium-Liquid Argon sampling calorimeter [1] was designed and built in the late 1980's for Run I of the Fermilab Tevatron $p\bar{p}$ collider. Some of the calorimeter electronics and external cabling were upgraded [2] for Run II of the collider to handle the smaller bunch spacing and higher luminosity.

The D0 calorimeter consisted of three cryostats, the central calorimeter (CC) which covers the pseudorapidity $|\eta| < 1.1$ and two end calorimeters (EC) which extend the coverage to $|\eta| < 4.5$, where $\eta = -\ln\tan(\theta/2)$. The CC is segmented into 8 layers while the ECs have 9 layers. The first four layers are used primarily to measure the energy of photons and electrons and are collectively called the electromagnetic (EM) calorimeter. The remaining layers (three or four fine hadronic (FH) and one course hadronic (CH) layers), along with the EM section, are used to measure the energy of hadrons. Most layers are segmented into $0.1 \times 0.1$ regions (cells) in $(\eta, \phi)$ space. The third layer of the EM calorimeter is segmented into $0.05 \times 0.05$ regions. Figure 1(left) shows a cross sectional *r* - *z* view of one quarter of the D0 calorimeter system.

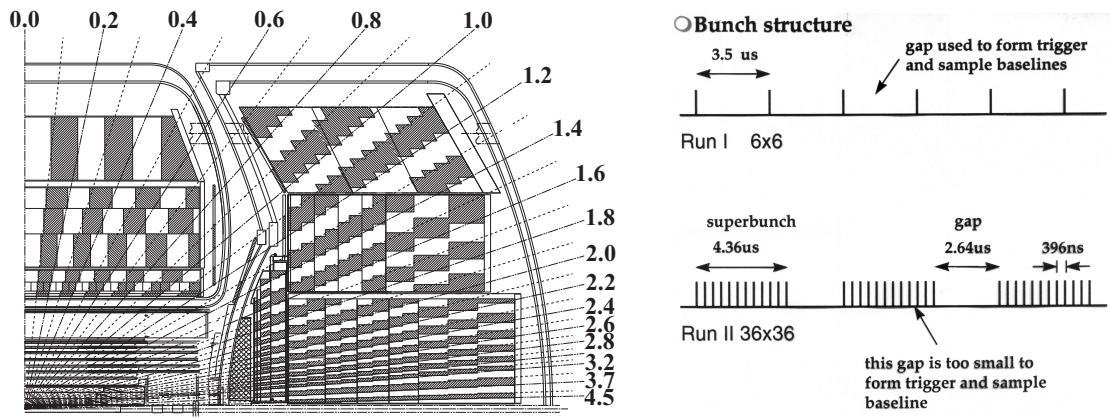

**Figure 1:** (*left*) Side view of one quarter of the D0 calorimeter system, showing segmentation and tower definitions. The lines extending from the center of the calorimeter denote the $\eta$ coverage of the cells and projected towers. The Run II solenoid and tracking detectors are also shown in the inner part of the detector. (*right*) Tevatron bunch structure for Run I and Run II.

## 2. Readout electronics and upgrade for Run II

The general design of the calorimeter readout electronics [3] remained essentially the same for both Run I and Run II but most of the on-detector electronics was updated to handle the higher crossing rate and luminosity for Run II. The calorimeter Level 1 (L1) trigger was also upgraded part way through Run II.

The $\approx$48,000 calorimeter signals exited through signal ports located near the top of the cryostats directly to charge sensitive preamps. The signals were routed through multilayer printed circuit boards which also reorganized the signals from module organization (many etas for a given layer and phi) to physics organization (all layers for a single eta and phi), greatly simplifying the trigger formation and cabling. The preamps and their associated low voltage power supplies were surrounded by the muon system including the iron magnets when the detector was in its operating configuration. Failures in the preamp system required at least one eight hour shift of machine downtime to open and close the iron for repairs, so the preamps were designed for high reliability and a redundant set of power supplies were installed and could be remotely swapped in case of a supply failure. The minimally shaped preamp signal was sent over $\approx$80 feet of cable to an area on the detector which was accessible in a short 15 minute machine access. There the signals were shaped separately for the precision readout and the trigger path. All 48 channels which constitute a calorimeter trigger tower (0.2 $\times$ 0.2 in $\eta, \phi$ space) were located on a single board. Local analog sums were created for the EM and hadronic layers separately. These trigger signals were then shipped over $\approx$300 feet of coax cables to the L1 calorimeter trigger system for digitization and trigger processing. The precision signal was stored in an analog memory to await an L1 decision. Two gain paths for each channel existed with an additional factor of eight amplification in the high gain path. When an L1 accept was received, a correlated double sample was made. The decision on whether to use the low or high gain path was then made. One analog signal along with a digital bit to indicate the gain path for each cell was shipped over $\approx$300 feet of cable to the precision readout digitization system. The signal was digitized to 12 bit precision, and the gain bit was used to produce a 15 bit value. All digitization was done outside the radiation area and was therefore accessible at any time.

The main difference affecting the readout electronics between Run I and Run II was the number and spacing of the beam crossings. All beam structure at the Tevatron was based on the 53.1046 MHz RF accelerator clock. It took 1113 clock cycles or about 21 $\mu$seconds for the beam bunches to travel around the ring. Figure 1(right) shows the beam structure for both Run I and Run II. In Run I there were six bunches with a beam spacing of $\approx$3.5 $\mu$seconds which allowed time to ship information to the trigger system and get back the results before the next crossing occurred. Only two sample and holds per channel were needed. The first sampled the baseline just before the crossing and the second sampled about 2.2 $\mu$seconds later when the signal from that crossing was at its maximum. A third sample and hold was used to store the difference of the two samples when an L1 accept was received, allowing for an essentially dead timeless readout of the calorimeter at the Run I L1 accept rates. In Run II there were thirty six bunches organized in three trains of twelve bunches. The closest spacing between bunches was 396 nanoseconds which required the use of a switched capacitor array (SCA) as the analog memory to hold forty one samples taken every 132 nanoseconds while waiting for the L1 decision. Because the SCA pedestals and noise were not adequate for 15 bits of dynamic range, we stored two gain paths in different SCA channels every 132 nanoseconds for each cell and chose the gain path to use only after the L1 accept was sent. The shaping of the signals was also changed to minimize the pileup while still using about two thirds of the total charge in the 2.3 mm Argon gap. In Run II there was a second level (L2) trigger system to reduce the rate of data sent to the farm of computers used to further filter the events sent to tape. This necessitated an additional L2 SCA storage element where the baseline subtracted value was held until an L2 confirmation was sent, before the signal was digitized. The analog cable drivers and digitization system for the precision readout were not upgraded for Run II.

## 3. Maintenance

To fully test spares for the detector we decided before Run I that we would build a realistic mockup of the readout path and use it to study and maintain the calorimeter electronics. It used a box of capacitors to simulate the readout cells of the detector, but used the spare parts from the real detector setup including the cables, crates and power supplies. We attempted to match the ground scheme as closely as possible, except that no other detector readout system shared the grounding. Since the detector was read out via four ports on each of the three cryostats which were essentially independent, we created a thirteenth setup which could handle up to one full port's worth of signals, but the system was typically only about half populated. The test setup was upgraded between Run I and Run II with new electronics just like the real detector and extensively used to test and burn-in the new electronics before installation on the detector. By the end of Run II enough of the spares in the test system had been used for failed parts in the real detector that the system was only capable of running half its original design capacity. The system could be run in two modes. The standalone mode used a PC to create triggers, fire pulsers and readout (at low rate) the system without needing the D0 trigger or DAQ system to be functional. The system also had the full set of trigger and DAQ connections, so could also run as an integrated part of the detector exactly like a thirteenth readout port.

After assembling the calorimeters in the three cryostats and welding them shut a total of 37 connections were found to be permanently bad out of the $\approx$48,000 channels. An addition 12 external connections were damaged during the Run II upgrade period outside the cryostat but in locations inaccessible without significant disassembly of other detector components. It was decided to add these to the permanently dead channel list. During Run I the typical number of additional channels marked bad for physics was less than 10, of which most were bad due to noise coming from the detector and therefore could not be fixed. Any electronics failure, mostly single channels, were typically fixed in less than a week during machine accesses. During Run II the number of noisy channels remained at about the same level but the typical number of bad electronics channel rose to about two dozen, primarily because of the higher complexity of the circuits and more types of failures which effected multiple channels. If the total number of channels flagged with bad electronics was greater than 48, the run was marked as POOR for calorimeter data and not used for many physics analyses. If more than 384 channels were flagged as bad electronics the run was marked BAD for calorimeter and no physics analysis which used the calorimeter data included the run in the analysis. That was always the case for a failed power supply since a single supply failure affected from eight hundred to four thousand channels.

## 4. Noise observed during the running of the D0 calorimeter

One of the most challenging issues in running a large liquid-argon calorimeter is maintenance of good understanding and control over any noise sources which might affect the performance of the calorimeter. Over the 20 years of running the D0 calorimeter a number of noise sources were identified. Once identified, some could eventually be eliminated while others needed to be efficiently flagged so that they would not affect the physics analysis, at the cost of some loss in efficiency.

The first noise observed early in Run I data was the "ring of fire" noise, named that because it effected all channels in a given depth in one of the two EC cryostats at the outer edge of the EM module in either layer 3 or 4. All 64 channels in a phi ring saw essentially the same amplitude coherent signal. Since EM energy was displayed in a red color in our event displays, it appeared as a red ring. It was identified as noise pickup on the high voltage (HV) distribution system inside the cryostat. Due to a poor design choice in how the HV was distributed to those EM layers, there was a significant and uniform capacitive coupling from the HV and each cell at the edge of the EC EM module. The rate of this noise was low but was never completely eliminated. Any events with this flag set were not used for physics.

The only other noise seen during Run I was the "hot cell" noise. It consisted of an isolated cell randomly located throughout the the detector. A typical isolated cell would become noisy for a few hours to a few months and then go quiet for a while. Sometimes it became noisy enough to affect the trigger rate, at which time the appropriate trigger tower was disabled until the cell got quiet or we had time to remove the associated summing resistor so that the rest of the tower could be re-enabled. The number and rate of this noise never significantly changed over the 20 years of operation and only effected a few channels in a given run. The cells were flagged on an event by event basis, unless the flagging rate for given cell exceeded a limit in which case the cell was marked bad for the entire run. A flagged cell had its energy divided by a billion, so was effectively turned off, but could still be accessed with special software for test purposes.

During Run II a few new types of noise were observed, the first of which was the "noon noise", named that because it usually started around lunch time and ended by dinner time. The noise affected a large fraction of the channels in the entire calorimeter. While the signal in a single cell was not that large, when summed over a large region of the calorimeter it completely compromised the jet and missing $E_T$ detector performance. It was eventually correlated with the use of a welding machine in the D0 assembly building. Once identified as the source we forbade the use of the welder while we were taking physics data or detector calibration runs. Because the detector was DC isolated from the building ground when built for Run I, except for a single safety ground with a large inductor to remove any high frequency noise, the size of the noise we saw was larger than expected. We spent significant time and effort during the next two machine maintenance periods identifying a number of accidental DC shorts to the building ground introduced between the end of Run I and the beginning of Run II while it was being upgraded, including a second safety ground without an inductor for high frequency filtering, all of which were fixed. While this reduced the size of the noise pickup, we kept the ban on welder use throughout the rest of Run II.

The next noise seen in Run II was noise associated with the Muon readout clock. Part way through Run II, the D0 silicon tracking system was upgraded to add an additional layer of silicon to improve its performance and help mitigate the effects of radiation damage to the existing silicon. The muon system used that machine shutdown period to slightly upgrade their readout system to allow a longer charge collection time to better measure late arriving hits. After the shutdown ended, the calorimeter pedestal widths were seen to have increased by about a factor of two. It was quickly associated with the muon system. It was then realized that this muon upgrade had the unfortunate side effect of moving their readout clock closer to the region of maximum sensitivity in the calorimeter readout system, creating a beat pattern with the baseline subtraction used in the calorimeter such that every pair of bunches saw the muon clock noise with exactly the opposite phase. If one sorted the data by bunch number, modulo four, you got the expected pedestal resolutions but the mean value took on two values separated by more than one sigma. Having understood this effect and its magnitude, we realized a smaller and more complicated pattern should have existed in the earlier Run II data. Sorting old pedestal data into the expected seven categories correlated to the previous muon clock rate showed a slight improvement in the RMS of the distribution and a small but measurable pedestal shift. The muon system then modified their readout clock to use the same clock frequency as the calorimeter and the noise was then completely removed by the correlated double sample. Occasionally during the rest of Run II, one or more of the muon chambers would lose phase with the machine clock. Since the data from those chambers were not usable for physics but did create noise in the calorimeter readout, the effected chamber electronics was powered down until the failed electronics could be replaced.

Another noise observed in Run II was the "purple haze" noise, named that because it effected essentially every channel in the CC calorimeter, so in the 2-D event display all you saw was the CH layer (purple) since almost every channel was above the zero suppression threshold in the CC calorimeter. It was first seen in the fall of 2005, but only in stores above a luminosity of about $60 \times 10^{30}$. The events were spectacular in that the occupancy was typically 80-100% and the minimum scalar $E_T$ seen in the events was $\approx$2 TeV and some events were above 12 TeV. Fortunately the rate was low ($\approx$50 events per hour), so we could take good physics data and simply flag the bad events to be skipped. We found that we could make a few similar events if we attempted to run the normal calorimeter physics triggers while we were ramping the HV. The following spring, during the shutdown to install the silicon detector upgrade we continued to study the effect by ramping the HV repeatedly until the effect suddenly disappeared. It returned in the fall of 2008. By that time the machine seldom ran below luminosity of $60 \times 10^{30}$ so was always present during physics runs. Fortunately during the shutdown for the silicon tracker upgrade the L1 calorimeter trigger was also upgraded and we now had access to the analog trigger sums even while taking good physics data and could configure a trigger which only ever fired on the purple haze noise. We soon isolated the source of the noise to a single HV power supply. We rewired it to 16 separate HV supplies, one for each of the 16 wires entering the cryostat, to locate the exact bad connection inside the cryostat. The wire turned out to be one of the two connections which supplied the same HV to both ends of the same set of resistive coat layers for redundancy and to minimize any voltage drop during periods of high current draw. By applying HV to one of the ends and shorting the other end we could now create the noise (without beam) on demand and at a reasonable rate to study it. Using a scope on the trigger sums and the special trigger, we studied what was happening. We saw the sign of the effect change depending on which end we applied HV to. The sign of the signal during normal running indicted which end was not directly connected to the detector, and only sparked when the voltage difference between the two ends was sufficient. The exact voltage necessary for sparking to occurred was difficult to measure because of the high resistivity of the resistive coating which made the rate of sparking go to zero as the voltage got close to the minimum value to spark. By allowing the HV line not directly connected to any resistive coat surfaces to float, the purple haze noise went away while keeping HV on the effected set of resistive coat surfaces. We ran that set of resistive coats connected only at one end for the rest of Run II.

During the next machine maintenance period we continued to study the system to get a better understanding of what was happening. We also hoped to understand why the noise disappeared for over two years and possibly make it go away again. The first step was to use a Time Domain Reflectometer (TDR) to locate the break inside the sealed CC cryostat. By comparing the signals on neighboring connected HV wires, the break was determined to be approximately at the location of the connector block on the end of the EM module where the HV wires make the final connection to the internal module cabling. Next we tried to determine the minimum voltage necessary to cause a spark across the break in the HV connection. At 30 volts there was no evidence of current or sparking. At 200 volts there was a high enough rate of sparking to clearly measure an average current and trigger on purple haze events. Somewhere near 100 volts there was evidence of sparks, but the average current observed was consistent with zero. Since we do not know the geometry for the break, only a crude estimate of the gap in the connection could be made at $\approx 6$ microns or less. We then tried to fix the HV connection by passing about 30 $\mu$amps of current through the break in the opposite direction from the current flow during normal physics running. After about 12 hours the current made a sudden jump and the noise disappeared. We retested the connection and there was clear DC current evident with only 30 Volts applied. The TDR measurements also confirmed the connection was now similar to its neighbors. Unfortunately about two weeks later the gap opened again.

Another noise observed in Run II was the "Spanish fan" noise, named that because it only effected channels at the very highest eta in the CC calorimeter so it looked in the 3-D event display like an open fan. Its rate was very low, about one per day and was only seen during times when the purple haze noise was present, but not in the same events. Presumably this noise is related to the HV spark which causes the purple haze noise but is either at the very beginning or the very end of the process when only the channels closest to the cryostat were involved.

The last noise observed in Run II was the "coherent" noise, named that because of its small but very coherent nature. It was seen in all parts of the calorimeter and was associated with the calorimeter readout itself. It only happened on L1 triggers which occurred while a previous L2 accepted trigger was being digitized. It was originally observed during normal physics running, but was then reproduced in the test setup and extensively studied. Only about one in six triggers taken during the L2 digitization time were effected, with no obvious correlation to when the trigger happened in the digitization cycle. The exact cause of this coherent noise was never identified, so these events were flagged and removed from all analyses which needed the calorimeter information. This amounted to between two and five percent efficiency loss depending on the L2 accept rate. The size of the effect was about two sigma for EM channels in the shift of the mean pedestal, either positive or negative.

## 5. Central Calorimeter Anomalous Currents

The D0 EC and CC calorimeters are very similar in construction. However the observed behavior of the HV bias currents in the two devices was very different. The ECs behaved as one might expect. When the bias voltage is turned on, the current rises to its equilibrium value nearly instantaneously and remains stable for long periods of time. Small jumps in the DC current level which last from hours to months before returning to it previous level are seen in some channels in both the EC and CC, and are attributed to movement of contaminants in the HV gaps. On the other hand, the currents in the CC takes several days to reach its equilibrium value and also show evidence of sharp (50 to 100 seconds) current discharges with a roughly constant frequency. Both the average current and the number and frequency of the discharges changed over time and/or integrated luminosity. Figure 2 shows the currents for the initial turn-on of the HV for a channel in both the CC and EC. Note the 2 days to reach equilibrium on the CC channel, while the EC reaches equilibrium within the 5 minute sampling interval.

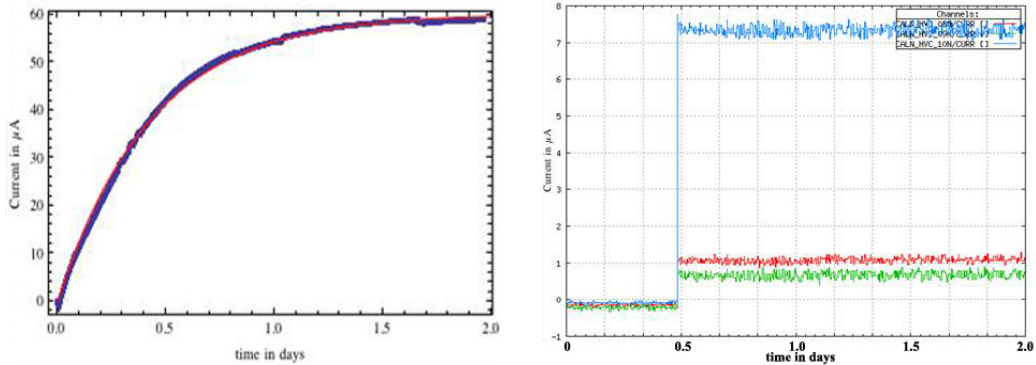

**Figure 2:** (*left*) Current for one CC HV channel as a function of time. (*right*) Current for three EC HV channel as a function of time.

So why are the two detectors so different? The most likely cause for the difference is that the uranium plates were processed differently for the CC and EC detectors during assembly. All the uranium plates came from the vendor with a rather thick layer of Uranium Dioxide ($UO_2$), which was unavoidable because Uranium oxidizes quickly in air. After seeing higher than expected currents in the test beam prototype modules, the EC assemblers decided to try removing some of the thicker oxide layer by use of a high pressure water jet in the hopes of reducing the anomalous currents, while the CC builders did not. The other half of the HV circuit is the resistive coat on the readout boards. This was a thin layer of a custom carbon/epoxy mix applied to the surface of the readout boards. The same application technique and resistance value ($\approx 150 \times 10^6$ Ohms per square) was used for both detectors. The shape of the readout boards is different between the CC and EC detectors, long and thin in the CC while squarer in the ECs. There is no obvious reason that the shape of the readout boards might change the current while the $UO_2$ layer is a clear candidate.

It has been known for over 60 years [4, 5] that a thin layer of high resistivity oxide on a metal surface in the presence of an electric field can cause positive ions to collect on the oxide layer. Under the proper conditions the field created by the ions can extract electrons from the metal and allow them to drift in the field. In the US this is referred to as Malter currents, a common problem in gaseous drift chambers. While the D0 calorimeter has liquid as the drift media, the same effect can occur (see Figure 3). Two distinct effects can occur. As positive ions collect on the insulating surface of the $UO_2$, the field across the thin oxide layer builds up, until electrons are extracted from the metal and are ejected into the gap. Some of the electrons neutralize the ions on the surface, while most drift across the gap and produce a continuous current. When the rate of neutralized argon ions is in balance with the new ions entering the region of Malter currents, the overall current is stable. If the field produced by the ions builds up sufficiently, the $UO_2$ insulator eventually breaks down and the current increases significantly, creating a Malter discharge. The discharge neutralizes a localized area of the insulator [6] as well as increasing the current across the gap. After the discharge the continuous current is lower than just before the discharge until the local positive ion density can build up enough that the discharges can occur again. Both Malter currents and discharges are seen in the CC calorimeter.

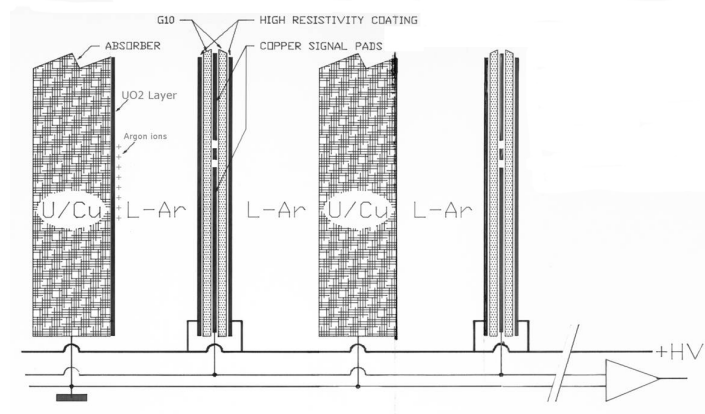

**Figure 3:** Basic D0 calorimeter readout cell structure.

No significant changes in the anomalous CC currents were noticed during Run I, but the HV current log is no longer available to carefully check if small changes occurred but were not noticed. In Run II, where both the length of the run and the integrated luminosity were higher a significant change in both the Malter currents and the Malter discharges was seen.

The Malter currents changed over the course of Run II. The average change was a factor of five. The change, viewed over the entire run, increased approximately linearly with integrated luminosity, implying some correlation to the total radiation dose. If that had also happened in Run I, then the effect would have been small enough to not be obvious given the large channel to channel differences. However two other parameters were observed to affect the change in Malter currents, independent of any change in integrated luminosity. First, if the HV is turned off for long enough to expect a change in Malter current levels, and then turned back on, the Malter current returns to the value it had at the time the HV was removed instead of the increasing as seen in other channels where the HV was left on. This was only tested after the accelerator was shut down, so no beam was present. This can be explained by assuming there is some contaminant that the HV is migrating to the $UO_2$ which is causing the Malter currents to increase. To make that consistent with the long timescale luminosity dependence, this contaminant must be produced by the radiation, and take a long time (of order months) to move to the $UO_2$ layer and affect the Malter currents. Second, the local average rate of rise in the Malter currents versus time without collisions was actually much higher than the rate observed with beam on. One possible explanation for this is that radiation damage causes two different effects. One happens during the radiation exposure and lowers the Malter currents, while the other is the contaminant discussed previously and takes time to affect the Malter currents and requires HV to be on. A candidate for this is radiation damage to the $UO_2$ creating grain boundaries and deep defects in the lattice [7, 8]. These competing effects would then make the increase in Malter currents higher during short accelerator down periods with HV on, then when the accelerator was running. Leaving the HV off during those times would delay slight the increase in currents but probably make little difference in the long term Malter currents seen on the CC calorimeter.

In the beginning of Run II only a few of the 32 HV channels had any Malter discharges. For those which had discharges, the discharges did not start when the HV was initially turned on, but began after a day or so when the Malter currents were beginning to stabilize. For a given HV supply the discharges were characterized by a unique frequency of discharge and amplitude. This is consistent with single localized area on the one of the approximately 64 different Uranium plate surfaces biased by that HV channel, and probably only a small region on that approximately 1.5 square meter surface which has exactly the correct configuration for discharges to be possible. By the end of Run II, essentially all HV channels saw Malter discharges and most had multiple discharges with different frequency/amplitude values. So whatever was happening to cause the Malter currents to increase over time also made new Malter discharge sites possible. There was a single example where it is clear that an existing Malter discharge site went away over time. After a Malter discharge, the level of Malter current dropped below the value seen just before the discharge and slowly returned to its previous value. During periods when the accelerator was off the frequency of a given Malter discharge was very stable. But when there were collisions the frequency of discharges changed. While they were faster at higher luminosity, the charge in frequency was small with respect to the change in current in the Argon gap, including both Malter and beam related currents.

## Acknowledgments

I wish to acknowledge the many members of the calorimeter operations group for their work in helping to maintain the detector electronics and to identify and classify the many sources and signatures of the different types of noise seen during the 20 years of operation of the D0 calorimeter. I wish to acknowledge Marvin Johnson for his help in studying and explaining the CC anomalous currents. I also thank the National Science Foundation for support.